\documentclass[12pt]{article}
\usepackage{graphicx,amssymb}
\usepackage{latexsym}
\usepackage{amsmath}
\usepackage{mathbbol}

\newtheorem{theorem}{Theorem}{\bf}{\it}
{\bf}{\rm}
{\bf}{\it}
{\bf}{\it}
\newtheorem{lemma}[theorem]{Lemma}{\bf}{\it}
          \def\C{{\mathcal C}}

          \def\H{{\mathcal H}}

          \def\del{\partial}
          \def\elzwo{\ell}

          \def\Halmos{\quad\hfill$\Box$}

            \def\jotzwo{\jmath}
          \def\integers{{\mathbb Z}}

          \def\naturals{{\mathbb N}}

          \def\reals{{\mathbb R}}

\title{Spin \& Statistics in Nonrelativistic\\ Quantum Mechanics, I}
\author{Bernd Kuckert\\
Korteweg-de Vries Instituut voor Wiskunde\\ Amsterdam, The
  Netherlands
\footnote{Present address: II. Institut f\"ur
    Theoretische Physik, Luruper Chaussee 149, 22761 Hamburg, Germany}}
\begin{document}
\maketitle

\begin{abstract}
A necessary and sufficient condition for Pauli's spin-statistics relation 
is given for nonrelativistic anyons, bosons, and fermions in two and 
three spatial dimensions.

For any point particle species in two spatial dimensions, denote 
by $J$ the total (i.e., 
spin plus orbital) angular momentum of a single particle, and let
$\jotzwo$ be the total angular momentum of the corresponding two-particle
system with respect to its center of mass. In three spatial dimensions,
write $J_z$ and $\jotzwo_z$ for the $z$-components of these vector operators.

In two spatial dimensions, the spin statistics connection holds 
if and only if there exists a unitary operator $U$ such that 
$\jotzwo=2UJU^*$. In three dimensions, the analogous relation cannot hold 
as it stands, but restricting it to an appropriately chosen 
subspace of the state space yields a sufficient and 
necessary condition for the spin-statistics connection.
\end{abstract}

\section{Introduction}

The best-known derivations of Pauli's spin-statistics connection
(which will, as usual, be called ``the'' spin-statistics connection in
what follows) have been found in quantum field theory, where various
proofs of increasing generality have been given over the decades.
Fierz and Pauli \cite{Fie39,Pau40} treated free fields, and L\"uders,
Zumino, and Burgoyne \cite{LZ58,Bur58} considered finite-component
general Wightman fields in 1+3 spacetime dimensions (see also
\cite{SW64}).  Similar results were obtained in the setting of
algebraic quantum field theory in 1+3 dimensions for both localizable
charges (Thm. 6.4 in \cite{DHR}) and topological charges \cite{BE85}.

Recently it has been found that for massive single-particle states of
quantum field theories, the spin-statistics connection can be derived
from the Unruh effect \cite{GL95} (cf. also \cite{FM91,BF02}) or from
a special form of PCT-symmetry \cite{Kuc95}, which follows from the
Unruh effect by an argument given in \cite{GL95}. Using an argument
given in \cite{BDFS}, one can further improve the result of
\cite{Kuc95}: the homogeneous part of the symmetry group does not need
to be the universal covering of the restricted Lorentz group; it
suffices to have the universal covering of the rotation group as
symmetry group \cite{Kuc03}. The strategy used in \cite{GL95,Kuc95}
has also led to spin-statistics theorems for anyons and plektons in
two spatial dimensions \cite{Lon97,Mun98}, conformal fields
\cite{GL96}, and quantum fields on curved spacetimes
\cite{GLRV,Ver01}.

Another approach to the spin-statistics connection is purely classical
(\cite{Tsch89,Tsch90}, cf. also \cite{Bal93}). It provides an
illustration of some crucial steps in the quantum field theory proofs
(cf. also the remarks made in \cite{Tsch91}) rather than a derivation
from first principles.

There have also been attempts to derive the spin-statistics connection
in the setting of nonrelativistic quantum mechanics.  But all
arguments suggested so far turned out to be based on too restrictive
assumptions, or they have been falsified by counterexamples (cf. the
discussions and references in \cite{DS98,DS,Wig99}). It has been shown
in \cite{BR00} that this also holds for the recent attempt by Berry
and Robbins \cite{Ber97}.

It is well known that quantum mechanics as such admits -- like quantum
field theory, see \cite{Str67} -- systems violating the
spin-statistics connection: the easiest examples are spinless
fermions, i.e., single-component wavefunctions that are antisymmetric
under particle exchange, also counterexamples with nonzero spin are
easy to find, and their second quantization is straightforward as well
(see, e.g., \cite{Wig00}). Each derivation of the spin-statistics
connection must rely on some additional assumption ruling out these
counterexamples.

In this Letter, we consider nonrelativistic anyons, bosons, and
fermions in two or three spatial dimensions and give a necessary and
sufficient condition for the connection between such a particle's spin
$\sigma$ and its statistics phase $\kappa\in S^1$, which Pauli discovered
to be 
\begin{equation}\label{ssc}
e^{2\pi\sigma i}=\kappa. 
\end{equation}
In three spatial dimensions, it has been shown that $\kappa\in\{\pm1\}$,
whereas in two dimensions, $\kappa$ can be any element of $S^1$
\cite{LD,LM}.

For a theory with nonabelian statistics, some additional structure
(e.g., a Markov trace) is needed to define a statistics phase and,
hence, to make the problem of finding a spin-statistics connection
well posed. The issue whether and how the subsequent argument can be
generalized to this case will be left open here.

In classical mechanics, the total angular momentum of two
indistinguishable particles with respect to their center of mass is
twice the angular momentum of each of the two with respect to the same
point of reference. Does this fact have a quantum counterpart?
Evidently, the observables to be compared with
each other will typically live in different Hilbert spaces, so any
analogous equality can, at most, be one up to a similarity
transformation by a unitary operator between these two Hilbert spaces,
i. e., up to unitary equivalence.

In the setting of two spatial dimensions, $J$ will denote the
total\footnote{In this paper, the word ``total'' is, as usual, to be
  read as ``spin plus orbital''.} angular momentum operator of a given
single particle in its one-particle space, and $\jotzwo$ will be the
total angular momentum of the corresponding two-particle system with
respect to its center of mass. 
It turns out that the spin-statistics
connection holds if and only if there is a unitary operator $U$ such
that $\jotzwo=2UJU^*$. 
This strong result is possible since the rotation
group $S^1$ is abelian, the consequence being that adding angular
momenta is analogous to the classical addition.

In three spatial dimensions, the situation is more involved, since the
rotation group and its universal covering are nonabelian, and
therefore the addition of angular momenta and spins is well known to
be more involved than in two dimensions. Denoting by $J_z$ the
$z$-component of the single particle total angular momentum and by
$\jotzwo_z$ the $z$-component of the two-particle system's total
angular momentum with respect to its center of mass, one finds that
the condition
\begin{equation}\label{bedingung}
\jotzwo_z=2UJ_z U^*
\end{equation}
cannot hold as it stands. Nevertheless one can look for some subspace
restricted to which the relation (\ref{bedingung}) is meaningful and
provides a sufficient and necessary criterion for the spin-statistics
connection.

To this end, the analysis of the three-dimensional case will be
confined to the Hilbert spaces $\H^{\uparrow}$ and
$\H^{\uparrow\uparrow}$ of all one-particle and two-particle states
where the $z$-components of all particle spins take their maximum
values.  Evidently these spaces are not invariant under most time
evolutions. This does, however, not affect the argument given below,
since it is purely kinematical: no Hamiltonian is specified, and the
free Hamiltonian, which one may use to specify the particle mass as a
further characteristic property of the particle, does commute with
spin.

Within $\H^{\uparrow\uparrow}$, denote by $\H_+$ and $\H_-$ the
eigenspaces of the $z$-parity operator $P_z: (x,y,z)\mapsto(x,y,-z)$
consisting of the functions in that are even or odd in the $z$-component,
respectively. It can be shown that Condition (\ref{bedingung}) holds
{\it either} when restricted to $\H_+$ {\it or} when restricted to
$\H_-$.  The spin-statistics connection is equivalent to the first
alternative, and its violation is equivalent to the second.

After specifying some setting and notation in Sect. \ref{setting}, the
two-dimensional case is discussed in Sect. \ref{two dimensions}. The
three-dimensional situation is discussed in Sect. \ref{three
  dimensions}, and some concluding remarks are made in Sect.
\ref{conclusion}.

\section{Setting and Notation}\label{setting}

The space of pure states of $n$ indistinguishable Bose or Fermi
particles in $s$ spatial dimensions can be defined by imposing either
symmetry or antisymmetry under particle exchange on a wave function in
$L^2(\reals^{sn})$.  Alternatively, one may first reduce the classical
configuration space by identifying indistinguishable configurations,
and then consider {\it all} wave functions on this space. In three
dimensions, it is a matter of taste which approach one wishes to use.
In two dimensions, however, particles whose statistics is neither
(para-) bosonic nor (para-) fermionic can occur, and these particles
can only be described in the second approach. For this reason, this
approach will be used in what follows.

Following Laidlaw and DeWitt \cite{LD}, the configuration space of $n$
distinguishable particles in $\reals^s$ is described by
the set $Y(n,s)$ of all $n$-tuples of $s$-vectors no two of which
coincide:
$$Y(n,s):=\{y=({\bf y}_1,\dots,{\bf y}_n)\in(\reals^s)^n:\,
{\bf y}_i\neq {\bf y}_j\,{\mbox{for}}\,i\neq j\}.$$
An action of the symmetric group $S_n$ on this space is defined by
$$\pi y:=({\bf y}_{\pi^{-1}(1)},\dots,{\bf y}_{\pi^{-1}(n)}),\quad\pi\in S_n.$$
The orbits of $S_n$ in $Y(n,s)$ yield the configuration space
$X(n,s):=Y(n,s)/S_n$ of $n$ indistinguishable particles. It is
straightforward to endow $X(n,s)$ with the structure of a pathwise
connected topological space for $s\geq2$ whose fundamental group is
$S_n$ for $s=3$ and the braid group $B_n$ for $s=2$. The fact that
there are only two scalar unitary representations of $S_n$ implies the
Bose-Fermi alternative for $s=3$ \cite{LD}, in two spatial dimensions,
arbitrary fractional statistics can occur as well \cite{LM}.

For $n=1$, one has $X(1,s)=\reals^s$, and a pure state of one particle
whose $z$-component of spin equals its maximum possible value $\sigma$
can be described by a one-component wave function on $\reals^s$ (all
possible other spinor components vanish). As usual, the $z$-component
of the orbital angular momentum is described by the self-adjoint
operator $L_z$, and the $z$ component of the total angular momentum
operator is $J_z=L_z+\sigma$.

For $n=2$, center of mass coordinates can be used to describe the
configuration space of two indistinguishable particles as the
cartesian product of $\reals^s$ and a relative coordinate space $\C$
\cite{LM}. In two spatial dimensions, $\C$ is the cone obtained (using
planar polar coordinates) from the half plane
$$\overline{H}:=\{(r,\varphi)\in\reals^2:\,r\geq0,-\pi/2\leq\varphi\leq\pi/2\}$$ 
by identifying $(r,-\pi/2)$ with
$(r,\pi/2)$ for each $r>0$ and by removing the origin at $r=0$. 
For $s=3$, $\C$ is obtained from the half space
$$\overline{H}:=\{(x,y,z)\in\reals^3:\,x\geq0\}$$
by identifying $(0,y,z)$ with $(0,-y,-z)$ for all
$(y,z)\in\reals^2$
and by removing the origin. 

In both two and three dimensions, $H$ will denote the interior of
$\overline{H}$.

\section{Two spatial dimensions}\label{two dimensions}

In two spatial dimensions, the pure-state space of two
indistinguishable particles both having spin $\sigma$ is canonically
isomorphic with the space $L^2(\reals^2\times\C)$. The wave functions
have one component only, since the universal covering group of the
rotation group is abelian, and since, as a consequence, its
irreducible representations are one-dimensional. Corresponding to the
interrelation between $\C$ and $H$ just discussed, any choice of the
above coordinates induces an isomorphism from $L^2(\reals^2\times\C)$
onto $L^2(\reals^2\times H)$ in a straightforward fashion.
Accordingly, the state space under consideration is
$$L^2(\reals^2 \times H,2d^2{\bf R}\,d^2{\bf r})\cong
L^2(\reals^2,d^2{\bf R})\otimes L^2(H,2d^2{\bf r}).$$
The orbital angular momentum operator in $L^2(\reals^2)\otimes L^2(H)$
with respect to the system's center of mass is of the form
$1\otimes\ell$, where $\ell$ is a self-adjoint operator in $L^2(H)$.
On the test functions with compact support in the interior of $H$, the
operator $\ell$ coincides with the familiar differential operator
$-i\del_\varphi$. But this {\it hermitian} differential operator
possesses many {\it self-adjoint} extensions $\elzwo$, which 
yield different unitaries $R:=e^{\pi i\,\elzwo}$. Since by
definition, the orbital angular momentum operator generates a
representation of the rotation group, one has $R^2=e^{2\pi i\,\elzwo}=1$.

If for some $\lambda\in\reals$, we define $\jotzwo:=\elzwo+\lambda$ as
the total angular momentum operator with respect to the center of
mass, then the statistics phase $\kappa\in S^1$ is related to
$\jotzwo$ by $\kappa=e^{\pi i\,\jotzwo}=Re^{\pi\lambda i}$.\footnote{
  As an aside, note that the exchange of two (indistinguishable) pairs
  of (indistinguishable) particles yields a braid diagram with {\it
    four} crossings, so the statistics phase of a two-particle system
  is $\kappa^4$; cf. also \cite{Fre90}. For arbitrary $n\in\naturals$,
  a little braid group diagrammatics shows that a rotation of $n$
  indistinguishable particles with respect to their center of mass is
  accompanied by a statistics phase $\kappa^{n(n-1)}$, whereas the
  $n$-particle system's statistics phase is $\kappa^{n^2}$, so one can
  assign a statistics phase $\kappa^{n(n-1)}$ to the relative motion
  and a phase $\kappa^n$ to the center of mass motion. This remark is
  redundant if $\kappa\in\{\pm1\}$ as in three dimensions, but since
  in two dimensions, $\kappa$ can take any value in $S^1$, it should
  be in place here.} It is to be emphasized that $\lambda$ is not
assumed to equal $2\sigma$ from the outset; it will, however, be found 
{\it as a result} that $\lambda-2\sigma$ is an even integer if Pauli's
spin-statistics connection holds.

Denoting the orbital angular momentum operator $-i\del_\varphi$ in the
single-particle space $L^2(\reals^2)$ by $L$,\footnote{To be more
  precise, the differential operator $-i\del_\varphi$ is well defined
  and essentially self-adjoint on the dense domain of test functions;
  $L$ denotes its (self-adjoint) closure.} the following theorem can be
shown.

\begin{theorem}
For $s=2$, the spin-statistics connection  
holds if and only if 
there
exists a unitary operator 
$U:L^2(\reals^2)\to L^2(H)$ 
such that
\begin{equation}\label{condition2}
\jotzwo=2(ULU^*+\sigma).
\end{equation}
\end{theorem}
{\it Proof.} Two lemmas will be used:
\begin{lemma}
For every integer $\nu$ with $(-1)^\nu R=1$,
there exists a 
unitary operator $U_\nu:L^2(\reals^2)\to L^2(H)$
with $2L=U_\nu^*\elzwo U_\nu+\nu$.
\end{lemma}
{\it Proof.} Denote the slit plane $\reals^{>0}\times(-\pi,\pi)$ by $H^2$, and
define
$$U_\nu\Psi(r,\varphi):=e^{-\nu\phi i}\Psi(r,2\varphi),\quad\Psi\in C_0^\infty(H^2),\quad 
(r,\varphi)\in H.$$
By Stone's theorem, it suffices to prove that
$$e^{-\vartheta i\cdot2L}\Psi=e^{-\vartheta i\cdot(U_\nu^*\elzwo
  U_n+\nu)}\Psi$$
for every $\Psi\in C_0^\infty(H^2)$ and every
$\vartheta\in\reals$. It turns out that this can be accomplished by
pointwise evaluating $e^{-\vartheta i\cdot2L}\Psi$ and $e^{-\vartheta
  i\cdot(U_\nu^*\elzwo U_n+\nu)}\Psi$ almost everywhere. Namely,
choose any $(r,\phi)\in H^2$ such that there exists a unique integer
$n_{\vartheta,\phi}$ with
$$-\pi<\phi-2\vartheta+2\pi n_{\vartheta,\phi}<\pi.$$
It is well known that
$e^{-\vartheta i\cdot2L}\Psi(r,\phi)=\Psi(r,\phi-2\vartheta+2\pi n_{\vartheta,\phi})$.

One concludes from this that
\begin{align*}
  e^{-\vartheta i(U_\nu^*\,\elzwo\, U_\nu+\nu)}&\Psi(r,\phi)
  =e^{-\nu\vartheta i}\,\,U_\nu^* e^{-\vartheta i\,\elzwo}U_\nu\Psi(r,\phi)
   =e^{\nu i(\phi/2-\vartheta)}e^{-\vartheta i\,\elzwo}U_\nu\Psi(r,\phi/2)\\
  &=e^{\nu i(\phi/2-\vartheta)}\quad\quad\quad e^{-\vartheta
    i\,\elzwo}
  \quad \quad\quad e^{-\nu(\phi/2)i}\,\Psi(r,\phi)\\
  &=e^{\nu
    i(\phi/2-\vartheta)}\,\,R^{n_{\vartheta,\phi}}\,\,e^{-i(\vartheta-\pi
    n_{\vartheta,\phi})\elzwo}\,
  e^{-\nu(\phi/2) i}\,\Psi(r,\phi)\\
  &=e^{\nu i(\phi/2-\vartheta)}\,\,(-1)^{\nu n_{\vartheta,\phi}}\,\,
  e^{-\nu i(\phi/2-\vartheta+\pi
    n_{\vartheta,\phi})}\,\Psi(r,\phi-2\vartheta+2\pi
  n_{\vartheta,\phi})\\
  &=\qquad\qquad\,\,(-1)^{\nu n_{\vartheta,\phi}}\qquad\quad\,(-1)^{\nu
    n_{\vartheta,\phi}}\,\Psi(r,\phi-2\vartheta+2\pi
  n_{\vartheta,\phi})\\
&=\Psi(r,\phi-2\vartheta+2\pi
  n_{\vartheta,\phi})=e^{-i\vartheta\cdot 2L}\Psi(r,\phi).
\end{align*}
Since this reasoning applies for almost all $(r,\phi)\in H^2$, this completes
the proof.\Halmos

\begin{lemma}\label{threecons2}
Any two of the following three conditions 
imply the third one:
\begin{quote}
(i) $e^{\pi i\,\jotzwo}=e^{2\pi\sigma i}$, i.e., Eq. (\ref{ssc}).\\
(ii) $\lambda\in 2\sigma+2\integers$.\\
(iii) $R=1$.
\end{quote}
\end{lemma}
{\it Proof.} If Condition (i) holds, then
$$1=e^{\pi\,i\elzwo}\cdot e^{\pi
  i\,(\lambda-2\sigma)}=Re^{\pi i(\lambda-2\sigma)}.$$
    It follows that
    Condition (i) implies [(ii)$\Leftrightarrow$(iii)].
It remains to show that [(ii)$\wedge$(iii)] implies (i):
$e^{\pi i\jotzwo}=e^{\pi i\elzwo}e^{\pi i(\lambda-2\sigma)}\cdot
e^{2\pi\sigma i}=e^{2\pi i\sigma}$.\Halmos

Finally, to prove Thm. 1, assume Condition (\ref{condition2}).
One then has
$$\kappa=e^{\pi i\,\jotzwo}=e^{\pi i\cdot2(ULU^*+\sigma)}
=e^{2\pi i\sigma}\,U\underbrace{e^{2\pi i\,L}}_{=1}U^*
=e^{2\pi i\sigma},$$
which is Eq. (\ref{ssc}).

Conversely, assume Eq. (\ref{ssc}) to hold. Then 
$e^{\pi i\,\jotzwo}=Re^{\pi i\lambda}=\pm e^{\pi\lambda i}=e^{2\pi \sigma i},$
so $\lambda-2\sigma\in\integers$.

If $\lambda-2\sigma$ is even, then Lemma 3 implies
$R=1=(-1)^{\lambda-2\sigma}$, so by Lemma 2, there exists a unitary intertwiner
between $2L$ and $\elzwo+\lambda-2\sigma$, which is Condition
(\ref{condition2}).

If $\lambda-2\sigma$ is odd, then Lemma 3 implies
$R=-1=(-1)^{\lambda-2\sigma}$.  Again, Lemma 2 implies
Condition (\ref{condition2}).\Halmos

\section{Three spatial dimensions}\label{three dimensions}

In three spatial dimensions, a relation analogous to Eq.
(\ref{condition2}) cannot hold on the whole Hilbert space. In order to
see this, first note that such a condition would, in particular, have
to hold for the one-particle- and two-particle states where the spins
of all particles are prepared at their highest possible values
$\sigma\in\frac{1}{2}\integers$ (one-particle states) and
$\lambda\in\integers$ (two-particle states), respectively. The
corresponding one-particle and two-particle spaces will be denoted by
$\H^\uparrow$ and $\H^{\uparrow\uparrow}$, respectively.

$\H^\uparrow$ is canonically isomorphic with $L^2(\reals^2)$, and
$\H^{\uparrow\uparrow}$ is canonically isomorphic with the space
$L^2(\reals^3\times\C)$, since all spinor components except one vanish.
Any choice of the above coordinates induces an isomorphism from
$L^2(\reals^3\times\C)$ onto $L^2(\reals^3\times H)$ in a
straightforward (while coordinate-dependent) fashion, so the state
space under consideration is
\begin{align*}
\H^{\uparrow\uparrow}&=L^2(\reals^3\times H,2d^3{\bf R}\,d^3{\bf
  r})\\
&\cong L^2(\reals^3,d^3{\bf R})\otimes L^2(H,2d^3{\bf r})=:
\H^{\uparrow\uparrow}_{\rm CM}\otimes\H^{\uparrow\uparrow}_{\rm
  rel}.
\end{align*}
The $z$-component of the orbital angular momentum operator in
$\H^{\uparrow\uparrow}_{\rm CM}\otimes\H^{\uparrow\uparrow}_{\rm rel}$
with respect to the system's center of mass is of the form
$1\otimes\ell_z$, where $\ell_z$ is a self-adjoint operator in
$\H^{\uparrow\uparrow}_{\rm rel}\cong L^2(H)$. When restricted to 
$C_0^\infty(H)$, the self-adjoint operator
$\ell_z$ coincides with the familiar hermitian differential operator
$-i\del_\varphi$. Since by definition, the orbital angular momentum
operator generates a representation of the group of rotations around
the $z$-axis in $L^2(H)$, the operator $R_z:=e^{\pi i\,\ell_z}$ 
is an involution, i.e., $R_z^2=e^{2\pi i\,\elzwo_z}=1$.

Now define $P_z\Psi(r,\varphi,z)=\Psi(r,\varphi,-z)$, and denote by
$\jotzwo_z:=\elzwo_z+\lambda$ the $z$-component of the total angular
momentum with respect to the center of mass.  Since
$$\kappa=P_z e^{\pi i\,\jotzwo_z}=e^{\pi\lambda i}P_z R_z,$$
(cf. the
remark made in Sect. 2 concerning the role of $\lambda$), and since
$R_z^2=P_z^2=\kappa^2=1$, one finds $R_z=\pm P_z$.

Next note that in this setting, the condition of Thm. 1 would still
lead to
$$1=U e^{2\pi i\, L_z}U^*=e^{\pi i(2UL_z U^*)}=e^{\pi i(\jotzwo_z-\lambda)}
\in S^1 R_z=S^1 P_z,$$
so it cannot hold as it stands. 

Defining
$\H_\pm:=\{\Psi\in\H^{\uparrow\uparrow}_{\rm rel}:\,P_z\Psi=\pm\Psi\}$, one
can prove
\begin{theorem}\label{ssc3}
  (i) For $s=3$, 
the spin-statistics connection (\ref{ssc}) holds if and only if 
there
  exists a unitary operator $U:\H^\uparrow\to
  \H^{\uparrow\uparrow}_{\rm rel}$ such that
\begin{equation}\label{condition31}
\jotzwo_z|_{\H_+}=2U(L_z+\sigma)U^*|_{\H_+}.
\end{equation}

(ii) For $s=3$, Eq. (\ref{ssc}) does not hold if and only if
there exists a unitary operator $U:\H^\uparrow\to
  \H^{\uparrow\uparrow}_{\rm rel}$ such that
\begin{equation}\label{condition32}
\jotzwo_z|_{\H_-}=2U(L_z+\sigma)U^*|_{\H_-}.
\end{equation}
\end{theorem}
{\it Proof.} The three-dimensional counterpart of Lemma 2 is
\begin{lemma}\label{uequiv3}
(i) For every integer $\nu$ with $R_z(-1)^{\nu}|_{\H_+}=1$,
there exists a unitary operator $U_\nu:L^2(\reals^3)\to
L^2(H)$ with 
$$(\elzwo_z+\nu)|_{\H_+}=2 U_\nu L_z U_\nu^*|_{\H_+}.$$

(ii) For every integer $\nu$ with $R_z(-1)^{\nu}|_{\H_-}=1$, there exists a 
unitary operator $U_\nu:L^2(\reals^3)\to L^2(H)$ with 
$$(\elzwo_z+\nu)|_{\H_-}=2 U_\nu L_z U_\nu^*|_{\H_-}.$$
\end{lemma}
{\it Proof.} In analogy to Lemma 2, define
$U_\nu\Psi(r,\varphi,z):=e^{-\nu\varphi i}\Psi(r,2\varphi,z)$ (using cylinder
coordinates), which, as above, intertwines between the hermitian
differential operators $-i\del_\varphi+\nu$ defined on the domain
$C_0^\infty(H)$ and $-2\del_\varphi'$ defined on the domain
$C_0^{\infty}(H^2)$, respectively. Using 
this operator in both cases, the proofs are completely analogous to that of
Lemma 2.\Halmos

The three-dimensional counterpart of Lemma 3 is
\begin{lemma}\label{threecons3}
(i) \quad Any two of the following three conditions imply the
third one.
\begin{quote}
(i.i) \quad $\kappa=P_z\,e^{\pi i\,\jotzwo_z}|_{\H_+}
=e^{\pi i\,\jotzwo_z}|_{\H_+}=e^{2\pi\sigma i}.$\\
(i.ii) \quad $\lambda\in 2\sigma+2\integers$.\\
(i.iii) \quad $R_z|_{\H_+}=P_z|_{\H_+}=1$.
\end{quote}

(ii) \quad Any two of the following three conditions imply the
third one.
\begin{quote}
(ii.i) \quad $\kappa=P_z\,e^{\pi i\,\jotzwo_z}|_{\H_-}=-e^{\pi i\,\jotzwo_z}|_{\H_-}=
-e^{2\pi\sigma i}$.\\
(ii.ii) \quad $\lambda\in 2\sigma+1+2\integers$.\\
(ii.iii) \quad $R_z|_{\H_-}=P_z|_{\H_-}=-1$.
\end{quote}
\end{lemma}
{\it Proof.} The proofs of the two statements are completely analogous
to the proof of Lemma \ref{threecons2} and will not be spelled out
here. \Halmos

Finally, to prove Thm. \ref{ssc3}, assume Condition (\ref{condition31}).
One then has
$$\kappa=\underbrace{P_z|_{\H_+}}_{=1}e^{\pi i\jotzwo_z}|_{\H_+}
=e^{\pi i\cdot2(UL_z U^*+\sigma)}|_{\H_+}
=e^{2\pi i\sigma}\,U\underbrace{e^{2\pi i\,L_z}}_{=1}U^*|_{\H_+}
=e^{2\pi i\sigma},$$
which is Eq. (\ref{ssc}).

Conversely, assume Eq. (\ref{ssc}). Then Condition (i.i) in Lemma
\ref{threecons3} holds. If $\lambda-2\sigma$ is even, then Lemma
\ref{threecons3}.i implies $R_z=1=(-1)^{\lambda-2\sigma}$.  Lemma
\ref{uequiv3} then implies that on $\H_+$, there exists a unitary
intertwiner between $2L_z$ and $\elzwo+\lambda-2\sigma$, whence
Condition (\ref{condition31}) follows.

If $\lambda-2\sigma$ is odd, then Lemma \ref{threecons3} implies
$R_z=-1=(-1)^{\lambda-2\sigma}$, and again, Lemma \ref{uequiv3} (i)
implies that on $\H_+$, there is a unitary intertwiner between $L_z$
and $\elzwo_z+\lambda-2\sigma$. This proves Statement (i).

The proof of Statement (ii) is completely analogous.
\Halmos

It is instructive to see what the conditions and statements of Theorem
4 look like when applied to the example of bound states of two
spinless Bose or Fermi particles interacting via some attractive
central potential. 

The parity of each bound state $\Psi\in\H^{\uparrow\uparrow}_{\rm
 rel}$ is $(-1)^l$, where $l$ is the azimuthal quantum number. 
By the indistinguishability of the two particles, only states
with either even or odd $l$ can occur, depending on whether the
particles are bosons or fermions, respectively. Evidently, the latter
violate the spin-statistics connection in the spinless case.

If the two particles are bosons, then $l$ must be even, and the
spin-statistics connection holds, so the additivity of angular momenta
must hold in $\H_+$ by Thm. 4. For each bound state $\Psi\in\H_+$,
the difference $l-m$ must be even as well, because
$P_z\Psi=(-1)^{l-m}\Psi$. It follows that $m$ is even and that
$U^*\Psi$ is an eigenvector of $L_z$ with the integer eigenvalue
$m/2$.

If on the other hand, the two particles are fermions, then only bound
states with odd $l$ occur,  and since the spin-statistics connection  is
violated, the additivity of angular momenta  holds in the space $\H_-$
by Thm  4. Reasoning as before,  one obtains  that $l-m$ is  odd for
bound states in $\H_-$ and that $m$, in  turn, is   even.  Again,
$U^*\Psi$  is  an eigenvector of   $L_z$ with  the  integer eigenvalue
$m/2$.

We find that the space $\H_+$ or $\H_-$ where the additivity condition
holds contains precisely the bound states in $\H_{\rm
  rel}^{\uparrow\uparrow}$ with even magnetic quantum numbers, as it
should be.

\section{Conclusion and Outlook}\label{conclusion}

The fact that in classical mechanics, the total angular momentum of a
system of two identical particles with respect to its center of mass
is twice that of each of the two particles, does, in parts, have a
quantum mechanical counterpart.

For nonrelativistic quantum mechanics in two spatial dimensions, it
turns out that this condition --- stated in terms of unitary
equivalence of the corresponding operators $\elzwo$ and $2L$ --- is
both sufficient and necessary for the spin-statistics connection.

In three spatial dimensions, the nonabelianness of
$SU(2)$ implies that the analysis has to be confined to the
$z$-components of the vector operators involved. It is
customary to confine the discussion to those states where all particle
spins have maximal $z$-components.  Within the space of these states,
the analogue to the two-dimensional additivity condition holds for
either the wave functions that are even in the $z$-component of their
relative coordinate or for the corresponding odd functions.

It turns out that the first alternative is equivalent to the
spin-statistics relation, whereas the second alternative is equivalent
to its violation.

The above results can be reformulated in a way that may be considered
as more natural, since the relative-coordinate space $\C$ does not
need to be ``cut open'' there in order to obtain the half space $H$
used above. This is currently being worked out together with Jens
Mund, and a corresponding joint paper will be published shortly
\cite{KM03}.

{\it Acknowledgements.} This work has been funded by the Stichting FOM
and the Deutsche Forschungsgemeinschaft.  I would like to thank D.
Arlt, D. Bahns, K. Fredenhagen, N. P.  Landsman, R. Lorenzen, J.-M.
Leinaas, J. Myrheim, M. Porrmann, K.-H. Rehren, T. Schlegelmilch, and
J. Zahn
for helpful discussions and questions. During the revision of this
manuscript, the discussions with Jens Mund have been a great help.

\end{document}